\documentclass[pre,aps,twocolumn,amsmath,amssymb]{revtex4-1}
\pdfoutput=1
\usepackage{hyperref}
\usepackage{graphicx}

\def\eqq#1{Eq.~(\ref{#1})}
\def\eq#1{(\ref{#1})}
\def\av#1{\left\langle #1 \right\rangle}
\def\f#1{Fig.~\ref{#1}}

\def\sec#1{Section~\ref{#1}}
\def\c#1{~\cite{#1}}

\def\deg{^\circ}

\def\kt{k_{\rm B}T}

\newcommand{\der}{{\rm  d}}

\def\dfdw{\ensuremath{\av{\partial f/\partial w}} }
\def\dfdphi{\ensuremath{\av{\partial f/\partial \phi_0}} }

\def\beq{\begin{equation}}
\def\eeq{\end{equation}}
\def\bea{\begin{eqnarray}}
\def\eea{\end{eqnarray}}

\begin{document}
\title{Irregular model DNA particles self-assemble into a regular structure}

\author{Zden\v ek Preisler$^1$}
\author{Barbara Sacc{\`a}$^2$}
\author{Stephen Whitelam$^1$}
\email[]{swhitelam@lbl.gov}
\affiliation{$^1$Molecular Foundry, Lawrence Berkeley National Laboratory, 1 Cyclotron Road, Berkeley, CA 94720, USA\\
$^2$Centre for Medical Biotechnology (ZMB) \& Centre for Nano Integration Duisburg-Essen (CENIDE), University of Duisburg-Essen, Universit\"{a}tstr. 2, 45117 Essen, Germany}

\begin{abstract}
DNA nanoparticles with three-fold coordination have been observed to self-assemble in experiment into a network equivalent to the hexagonal (6.6.6) tiling, and a network equivalent to the 4.8.8 Archimedean tiling. Both networks are built from a single type of vertex. Here we use analytic theory and equilibrium and dynamic simulation to show that a model particle, whose rotational properties lie between those those of the vertices of the 6.6.6 and 4.8.8 networks, can self-assemble into a network built from {\em three} types of vertex. Important in forming this network is the ability of the particle to rotate when bound, thereby allowing the formation of more than one type of binding motif. The network in question is equivalent to a false tiling, a periodic structure built from irregular polygons, and possesses 40 particles in its unit cell. The emergence of this complex structure, whose symmetry properties are not obviously related to those of its constituent particles, highlights the potential for creating new structures from simple variants of existing nanoparticles.
\end{abstract}

\maketitle

\section{Introduction}

DNA nanotechnology has provided a large number of nanoparticles that are able to self-assemble into interesting structures\c{seeman1994dna, seeman2010nanomaterials}. Usually, the symmetry of the structure is related in a simple way to the symmetry of the constituent nanoparticle. For example, 3-pointed DNA `star' particles, whose sticky arms lie 120$\deg$ apart, can self-assemble into a network equivalent to the hexagonal tiling\c{doi:10.1021/ja0541938} (this tiling is also designated 6.6.6, meaning that three regular 6-gons encircle each vertex\c{grunbaum1977tilings}). In \f{fig1} we show a picture of a network, built from model particles, equivalent to the 6.6.6 tiling (drawing lines between the centers of bound particles produces the picture of the tiling). The angle between sticky patches, 120$\deg$, is equal to the internal angle of the hexagon, and so the symmetry of the particle is clearly reflected in the symmetry of the assembly. Three-fold-coordinated DNA particles can also self-assemble into a network equivalent to the 4.8.8 tiling\c{grunbaum1977tilings,ANIE:ANIE201601944} (meaning that a square and two octagons encircle each vertex). The model picture in \f{fig1} shows such a network, built from a particle whose sticky arms lie 90$\deg$ and 135$\deg$ degrees apart. The angles 90$\deg$ and 135$\deg$ are equal to the internal angles of the square and the octagon, making clear the connection between the particle and the assembly. Notably, the sticky patches of the particle must be chemically specific, in the sense that only two types of complementary interaction are possible\c{ANIE:ANIE201601944,whitelam2016minimal}. In the absence of chemical specificity, if all particles stick to all other particles, then there exist too many competing structures to allow self-assembly of the 4.8.8 tiling. In this sense the 4.8.8 tiling is more complex than the 6.6.6 tiling.\\
\begin{figure}[t] 
   \centering
   \includegraphics[width=0.9\linewidth]{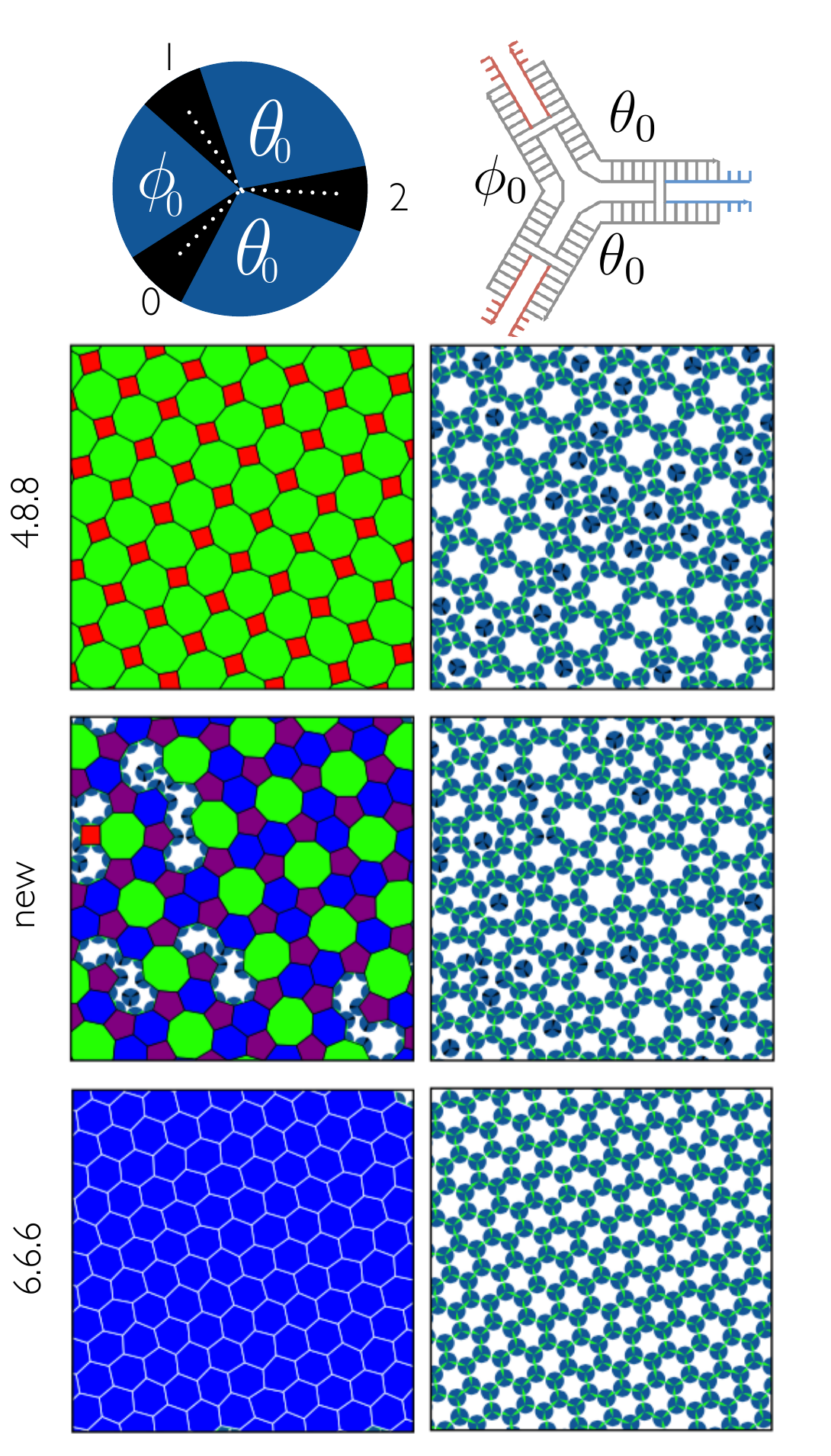} 
	\caption{Top: Geometry of model particles, together with a schematic of the type of DNA nanoparticle they are intended to represent (see e.g. Ref.\c{ANIE:ANIE201601944}). The angles $\phi_0$ and $\theta_0$ are the angles separating patch bisectors. Only the patch combinations 0:1 and 2:2 are sticky. Bottom: snapshots of the networks self-assembled when $\phi_0$ takes different values. From top to bottom: For $\phi_0\approx 90\deg$ (here $96 \deg$) the network equivalent to the 4.8.8 tiling; for $100\deg \lesssim \phi_0 \lesssim 110\deg$ (here $104 \deg$), the false tiling, labeled `new' in figures; and for $\phi_0 \approx 120\deg$ (here $116 \deg$) the 6.6.6 tiling.  The parameter $w=10\deg$ in all cases.}
   \label{fig1}
\end{figure}
In this paper we describe an example of self-assembly in which the assembled structure is not related in an obvious way to the symmetry of its constituent particles. We use analytic theory and simulation to study a patchy-particle model \cite{kern2003fluid,zhang2004self,doye2007condensed,wilber2007reversible,bianchi2008theoretical,sciortino2010,romano2012,romano2011,vissers2013,preisler2016} of 3-pointed DNA nanoparticles. When the rotational properties of this particle lie between those of the constituents of the 4.8.8 and 6.6.6 networks, we observe the assembly of a network more complex than either. A section of this new network is shown in \f{fig1}. The network possesses three types of vertex, around which, in clockwise order, sit 1) a pentagon, a hexagon, and an octagon; 2) a pentagon, an octagon, and a hexagon; and 3) a pentagon and two hexagons. These polygons are irregular -- the tiling is `false'\c{grunbaum1977tilings} -- because the sum of angles of regular polygons of these types is not 360$\deg$. Although these vertices and polygons are irregular, the network is periodic over a distance of several unit cells. Just as for the 4.8.8 network, assembly of this false tiling requires chemical specificity. However, in contrast to the 4.8.8 or 6.6.6 networks, the interactions of the constituent particles must be flexible to a substantial degree, to allow the adoption of more than one type of binding motif. Neither motif is exactly commensurate with the rotational properties of the particle itself. 

In what follows we describe the dynamic simulations used to discover this structure (\sec{dynamic}), and the analytic (\sec{analytic}) and equilibrium (\sec{equilibrium}) simulation methods that suggest that the structure is a thermodynamically stable phase. We conclude in \sec{conclusions}.

\section{In which dynamic simulations reveal the self-assembly of a peculiar structure}
\label{dynamic}

\begin{figure*}[t] 
   \centering
   \includegraphics[width=\linewidth]{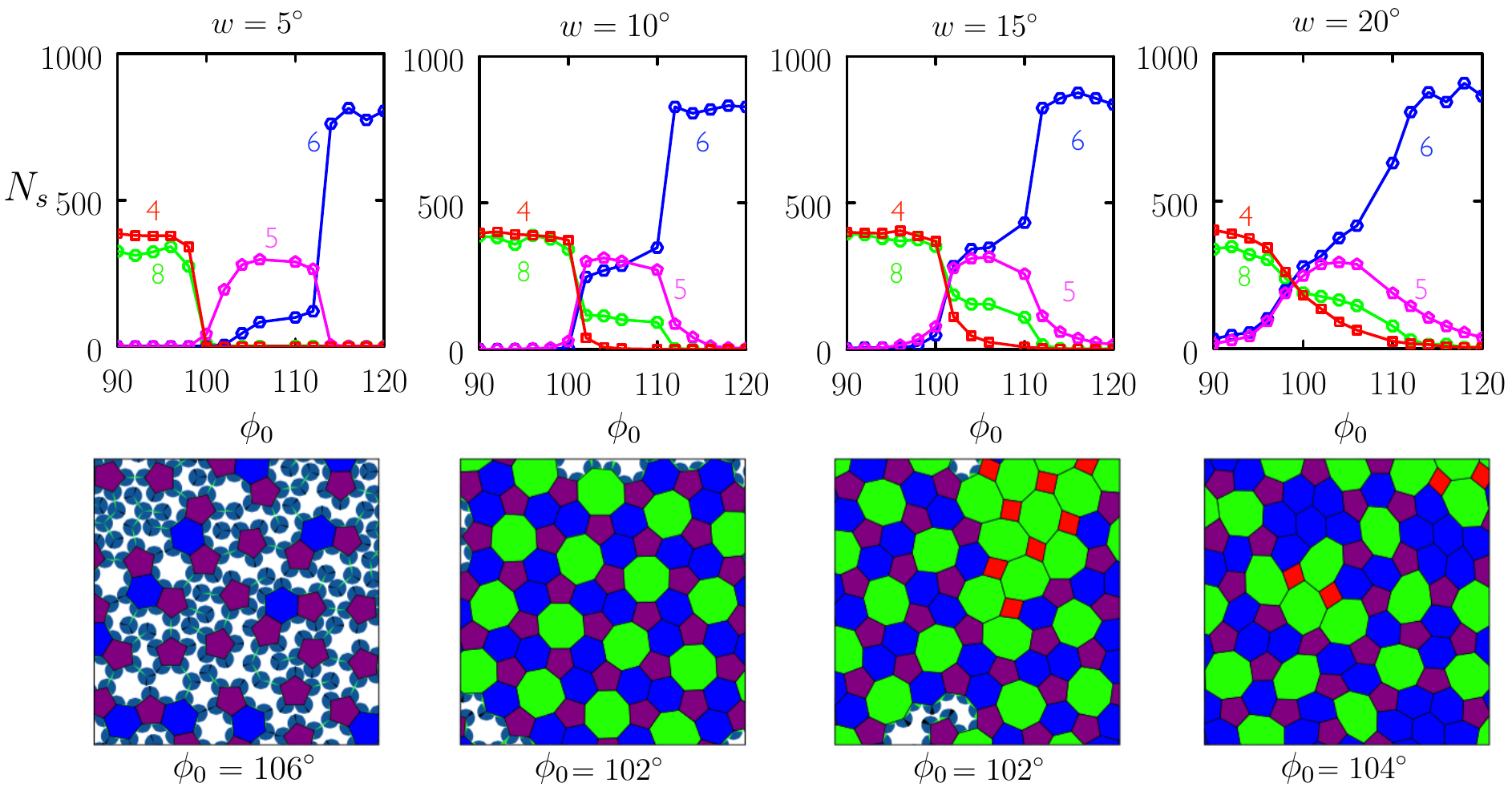} 
   \caption{Largest number of convex $s$-gons, $N_s$ formed by networks over the course of self-assembly simulations, for a range of patch widths $2w$ and small patch-bisector angles $\phi_0$. Shown below are snapshots taken at the indicated values of $w$ and $\phi_0$.}
   \label{fig2}
\end{figure*}

Consider a patchy-particle simulation model similar to those of Refs.\c{whitelam2015emergent,whitelam2016minimal}, sketched in \f{fig1}. We take hard discs of diameter $a$ that attract each other via patches of opening angle $2w$. The larger is $w$, the more the particle can rotate when bound, i.e. $w$ can be considered a proxy for binding flexibility. Patch bisectors are separated by angles $\phi_0$ and $\theta_0 = \pi - \phi_0/2$; we shall focus on the interval $90\deg \leq \phi_0 \leq 120\deg$. Particle interactions are of the Kern-Frenkel\c{kern2003fluid} type, and are chemically selective. In units of $\kt$, particles experience a pairwise attraction of energy $-\epsilon$ if 1) two disc centers lie within a distance $11\sigma/10$, where $\sigma$ is a particle diameter (an arbitrary choice intended only make the range of interaction short with respect to the particle diameter); if 2) the line joining the two disc centers cuts through one patch on each disc; and if 3) the contacting patches are of type 0 and 1 or 2 and 2.  Engaged patches are shown green in figures, and free patches are shown black. We sometimes draw, on top of networks, the convex polygons that result from joining the centers of bound discs. The simulation protocol we used is designed to mimic the deposition and assembly of molecules on surfaces. We work in the grand-canonical ensemble. We start with an empty substrate and allow particles to appear and disappear on it using grand-canonical Monte Carlo moves\c{frenkel1996understanding}. We allow particles to move on the substrate using the virtual-move Monte Carlo algorithm described in the appendix of Ref.\c{whitelam2009role}. We started the simulation with a small value of $\epsilon$ ($\approx 1$) and a chemical potential chosen so that the substrate is sparsely occupied by particles, and we `cooled' the system slowly by increasing $\epsilon$ by a value of $\approx 0.1$ every million Monte Carlo steps. After some time we observe the formation of a network structure, either ordered or disordered.

To identify networks of interest, we recorded the largest number $N_s$ of convex polygons of type $s$ ($s$-gons) seen over the course of the cooling simulation (recall that polygons are drawn on top of networks by joining the centers of bound discs). Results for a range of choices of patch width $2w$ and small patch-bisector angle $\phi_0$ are shown in \f{fig2}. When $\phi_0 \approx 90\deg$ we observe assembly of a network equivalent to the 4.8.8 Archimedean tiling\c{whitelam2016minimal}, signaled on the plots by the roughly equal numbers of 4-gons and 8-gons. When $\phi_0 \approx 120\deg$ we observe assembly of a network equivalent to the 6.6.6 tiling. Note that this network assembles with or without patch chemical specificity. For intermediate values of $\phi_0$ we observe networks that harbor a mixture of convex $n$-gons, including 8-gons, 5-gons and 6-gons. Some of those structures are shown in the bottom panel of the plot. For values of $w$ between about 10$\deg$ and about 20$\deg$, and for values of $\phi_0$ between about 100$\deg$ and 110$\deg$, we observe self-assembly of the aforementioned false tiling, signaled by the roughly 1:2:2 ratio of 8-gons:6-gons:5-gons. This tiling assembles in small patches of a few unit cells, unlike the 4.8.8 and 6.6.6 tilings, which typically occupy the whole simulation box. For larger values of $w$ the false tiling coexists with the 4.8.8 tiling (for $\phi_0\approx 100 \deg$) -- see the third snapshot in \f{fig2} from the left -- or the 6.6.6 tiling (for $\phi_0\approx 110 \deg$). 

The following two sections show that false tiling, in the parameter regime we see it assemble, is either thermodynamically stable or comparable in free energy to the 6.6.6 and 4.8.8 tilings.

\section{In which analytic calculation suggests that the false tiling is thermodynamically stable}
\label{analytic}
\begin{figure}[b] 
   \centering
   \includegraphics[width=\linewidth]{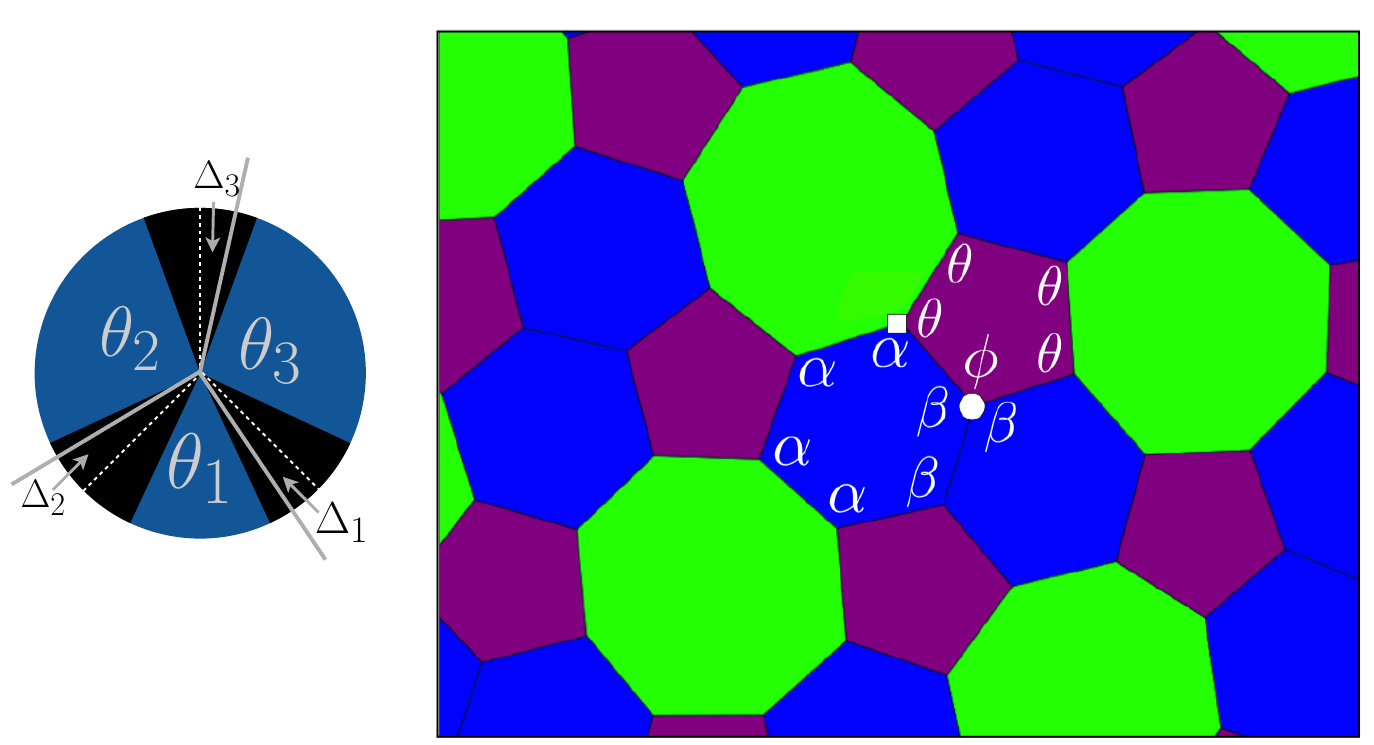} 
   \caption{Graphical construction for the analytic argument described in \sec{analytic}. The angles $\theta_i$ are the internal angles of three polygons that meet at the vertex of a network, and the angles $\Delta_i$ describe the mismatch between these angles and the angles $\phi_0$ and $\theta_0$ between patch bisectors.. As described in the text, the deflection angles $\Delta_i$ must satisfy certain criteria for a disc to serve as such a vertex, and the associated freedom defines the particle's rotational free energy. Right: Geometry of the false tiling.}
   \label{fig3}
\end{figure}
In this section we show that a simple analytic estimate indicates that the false tiling is lower in rotational free energy than either the 6.6.6 or 4.8.8 tilings, in the region of parameter space in which we see it self-assemble. The argument makes use of the geometric construction shown in \f{fig3}. The argument is mean-field in nature: we imagine a single disc sitting at a single vertex, and we calculate the rotational entropy the disc possesses if it is bound, i.e. if the edges of the network emanating from the vertex cut thorough all three of its patches. We imagine the particle's position to be fixed to the center of the vertex, and so we ignore vibrational entropy. We also ignore the configurational entropy associated with rearrangements of the network. Comparison with the simulations of \sec{equilibrium} indicates that these approximations are reasonable, i.e. that the properties of networks are indeed dominated by rotational entropy.

If a disc (whose patch bisector angles are $\phi_0$ and $\theta_0=\pi-\phi_0/2$) sits at a vertex between polygons of internal angles $\theta_1$, $\theta_2$ and $\theta_3$, with the lines separating polygons offset from the patch bisectors by angles $\Delta_1$, $\Delta_2$, and $\Delta_3$, then 
\bea
\theta_1 &=& \phi_0 - \Delta_1+\Delta_2; \nonumber \\
\theta_2 &=& \theta_0 - \Delta_2+\Delta_3; \quad {\rm and}\nonumber \\
\theta_3 &=& \theta_0 - \Delta_3+\Delta_1.
\eea
It is convenient to solve the above equations for $\Delta_2$ and $\Delta_3$ in terms of $\Delta_1$, giving
\bea
\label{subs}
\Delta_2 &=& \theta_1 -\phi_0+\Delta_1;   \quad {\rm and}\nonumber\\
\Delta_3 &=& \pi - \phi_0/2-\theta_3+\Delta_1,
\eea
with $\theta_2=2\pi - \theta_1-\theta_3$. 

We want to evaluate the rotational partition function for a bound disc,
\beq
\label{zed}
Z= \int_{-\pi}^\pi {\rm d} \Delta_1 \prod_{i=1}^3 \Theta(w-\Delta_i)\Theta(w+\Delta_i),
\eeq
in which $\Theta(x)=1$ if $x>0$ and $\Theta(x)=0$ otherwise. These $\Theta$ functions are all 1 if each grey line in \f{fig3} cuts through a black (sticky) patch. The largest possible value of $Z$ is $2 w$, which occurs when the polygon angles $(\theta_1,\theta_2,\theta_3)$ equal the patch-bisector angles $(\phi_0,\pi-\phi_0/2,\pi-\phi_0/2)$.

With all arguments in \eq{zed} written out we have
\bea
Z(\phi_0,w;\theta_1,\theta_3) &=& \int_{-\pi}^\pi {\rm d} \Delta_1 \Theta(w-\Delta_1)\Theta(w+\Delta_1)\nonumber \\
&\times& \Theta(w-\theta_1 +\phi_0-\Delta_1)\nonumber \\
&\times&\Theta(w+\theta_1 -\phi_0+\Delta_1) \nonumber \\
&\times& \Theta(w-\pi + \phi_0/2+\theta_3-\Delta_1)\nonumber \\
&\times&\Theta(w+\pi - \phi_0/2-\theta_3+\Delta_1),
\eea
with $\theta_2=2\pi - \theta_1-\theta_3$.

For the rotational free energy of the 6.6.6 tiling we have $\theta_1=\theta_3=2\pi/3$, and so, in units of $\kt$,
\beq
\label{f666}
f_{666}(\phi_0,w) = -\ln Z(\phi_0,w;2\pi/3,2\pi/3).
\eeq
For the rotational free energy of the 4.8.8 tiling we have
\beq
\label{f488}
f_{488}(\phi_0,w) = -\ln Z(\phi_0,w;\pi/2,3 \pi/4).
\eeq
\begin{figure}[t] 
   \centering
   \includegraphics[width=0.8\linewidth]{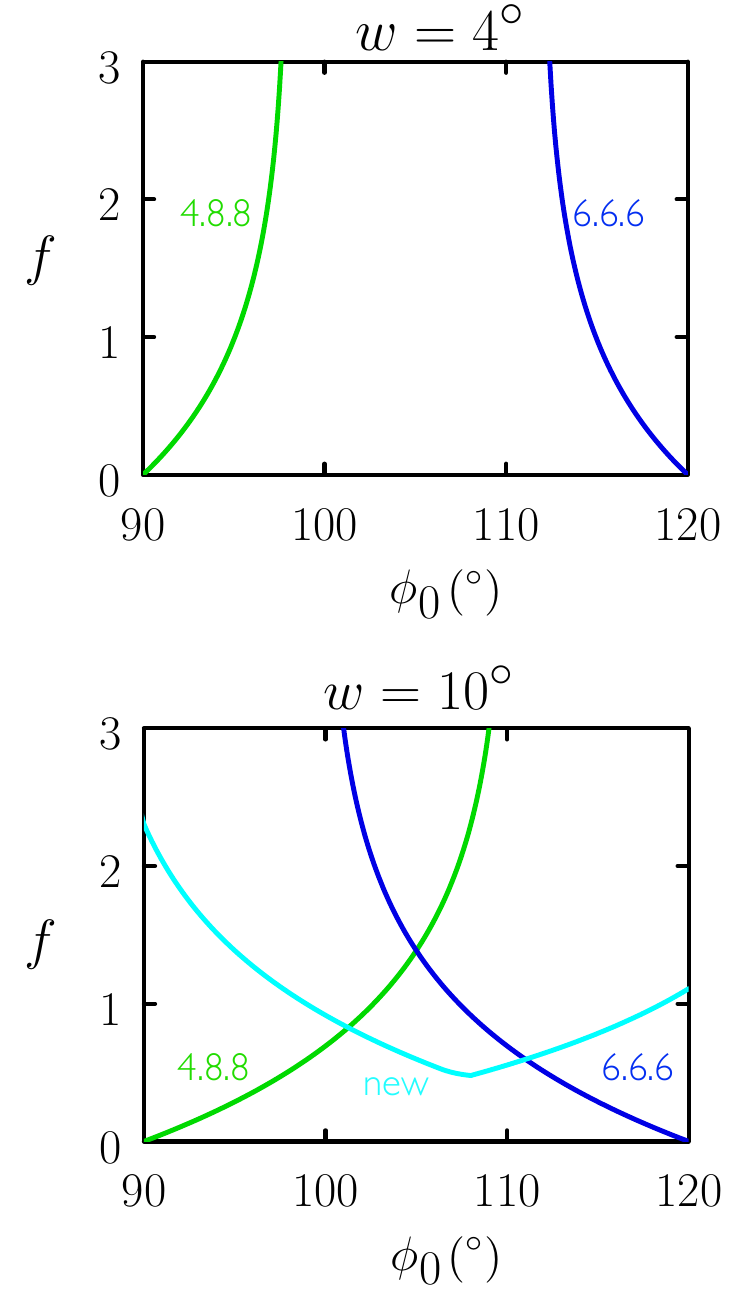} 
   \caption{Rotational free energy $f$ (in units of $\kt$) of discs with small patch-bisector angle $\phi_0$ that serve as the vertices of the indicated tilings. Lines are obtained from numerical evaluation of Equations \eq{f666}, \eq{f488}, and \eq{fnew2}. We have subtracted from each line the smallest possible value of $f$, namely $\ln 2w$, so that 0 is the origin of free energy.}
   \label{fig4}
\end{figure}
To compute the rotational free energy of the false tiling we refer to \f{fig3}(b). There are two vertex types labeled in this picture, $\circ$ (5.6.6) and $\square$ (5.8.6 or 5.6.8). To work out the number of vertices of each type, note that every octagon has a half-share (with another octagon) of 4 pentagons, and a half-share (with another octagon) of 4 hexagons. Thus the ratio of octagons:hexagons:pentagons is 1:2:2. Each $n$-gon has a one-third share of $n$ vertices, and so the total number of vertices in this fundamental unit is $(8+2 \times 6+2 \times 5)/3 = 10$. Each octagon has only $\square$ vertices; each hexagon has 4 $\square$ vertices; and each pentagon has 4 $\square$ vertices. Thus in the fundamental unit of the pattern we have $(8 + 2 \times 4 + 2\times 4)/3=8 $ vertices of type $\square$, and so the probability that a given vertex is of type $\square$ is $8/10=4/5$. Thus we take
\bea
\label{fnew}
f_{{\rm new}}(\phi_0,w) &=& -\frac{4}{5}\ln Z(\phi_0,w;\theta,\alpha)\nonumber \\
&-&\frac{1}{5}\ln Z(\phi_0,w;\phi,\beta).
\eea
To determine the angles $\theta, \phi, \alpha$ and $\beta$ characterizing the false tiling we again refer to \f{fig3}. Note that the sum of the internal angles of an $n$-gon is $(n-2)\pi$, whether or not the $n$-gon is regular. Thus the sum of the pentagon's angles is $4 \theta + \phi = 3 \pi$, and the sum of the hexagon's angles is $4 \alpha + 2 \beta = 4 \pi$. At the vertex $\circ$ we have $2 \beta + \phi = 2 \pi$. As a simplifying assumption we take the octagon to be regular, with each internal angle being $3 \pi/4$. Then we have $\theta + \alpha + 3\pi/4= 2\pi$. We can therefore determine all the angles of the tiling in terms of one parameter, $\phi$. We then take the free energy of the tiling to be the minimum of \eqq{fnew} as we vary this parameter, i.e.
\begin{widetext}
\beq
\label{fnew2}
f_{{\rm new}}(\phi_0,w) = -\min_\phi \left[\frac{4}{5}\ln Z\left(\phi_0,w;\frac{3\pi-\phi}{4},\frac{\pi}{2}+\frac{\phi}{4}\right)+\frac{1}{5}\ln Z\left(\phi_0,w;\phi,\pi-\frac{\phi}{2}\right)\right].
\eeq
\end{widetext}
In \f{fig4} we show the rotational free energies, as a function of the angle $\phi_0$, for discs that serve as the vertices of the three structures. We show plots for patch half-angle $w =4\deg$ and $w=10\deg$. For the narrower patch only the 4.8.8 and 6.6.6 tilings are viable, close to the ideal values $\phi_0=\pi/2$ and $\phi_0=2\pi/3$. For the wider patch, however, the false tiling is also viable, and is lower in free energy than the 4.8.8 and 6.6.6 tilings for intermediate values of $\phi_0$. 

In \f{fig5} we identify where, in the space of $\phi_0$ and $w$, each of the three phases is lowest in free energy. In the region marked `coex' the neighboring phases have rotational free energies within $\kt/10$ of each other -- i.e. rotational free energy does not strongly discriminate between the phases -- and so other factors (such as vibrational free energy, ignored by this simple calculation) will be important in determining phase behavior. Thus in the region in which we see self-assembly of the false tiling in simulations (\sec{dynamic}), $10\deg \lesssim w \lesssim 20\deg$ and $100\deg \lesssim \phi_0 \lesssim 110\deg$, the simple analytic argument of this section predicts that it is thermodynamically stable with respect to the 6.6.6 and 4.8.8 tilings. In the following section we show that free-energy simulations confirm this expectation.
\begin{figure}[t!] 
   \centering
   \includegraphics[width=\linewidth]{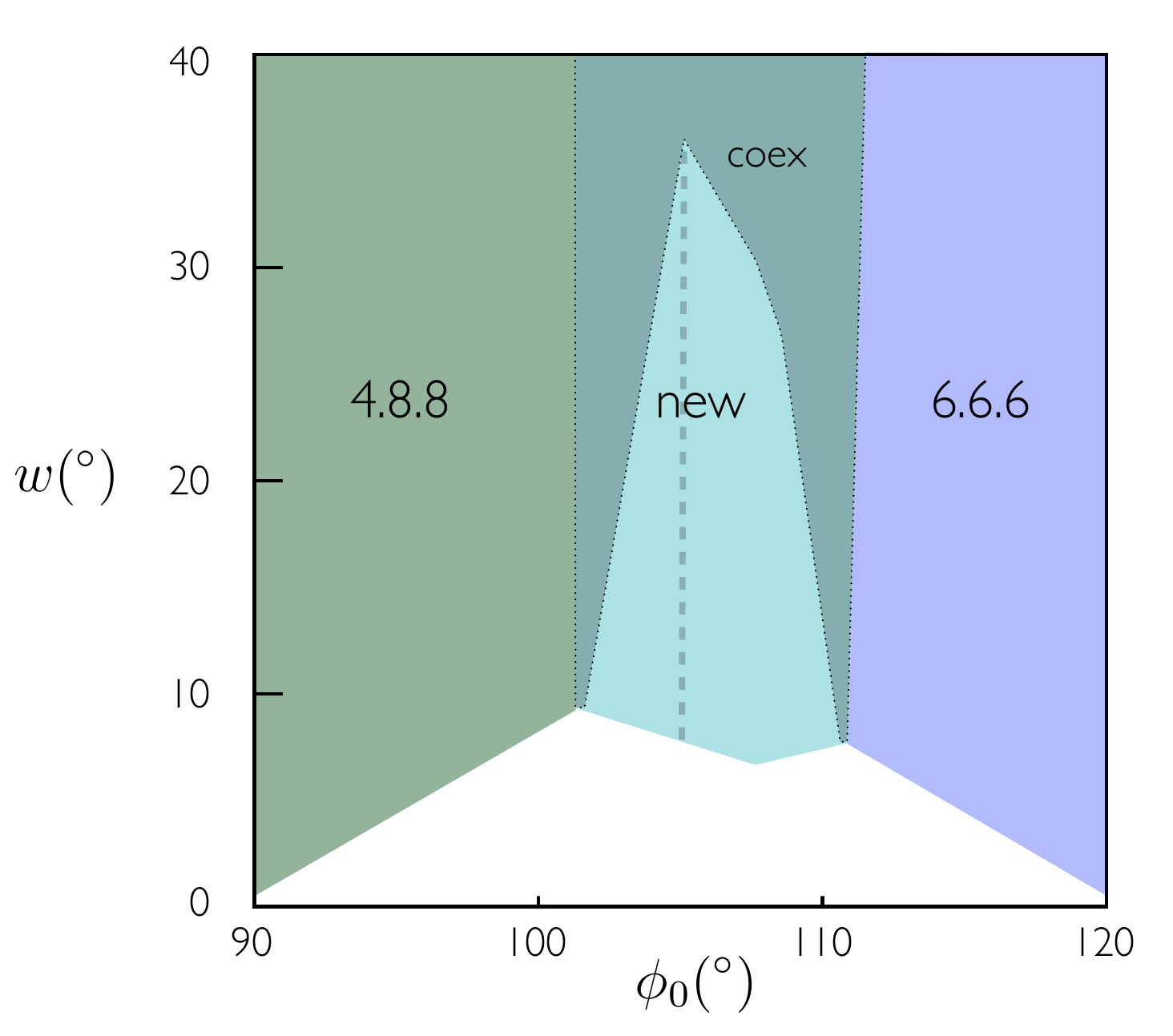} 
   \caption{Regions of parameter space in which the indicated phase is lowest in rotational free energy, according to the equations \eq{f666}, \eq{f488}, and \eq{fnew2}. We show only phases whose rotational free energies lie within $\kt$ of $\ln 2 w$: thus in the white region all phases are strained (this strain can be inferred from simulation: see e.g. the left snapshot in \f{fig2}, in which no ordered network is visible). In the gray region (labeled `coex') the neighboring phases have rotational free energies that lie within $\kt/10$ of each other. The gray dashed line indicates the points at which the 4.8.8 and 6.6.6 networks are equal in rotational free energy.}
   \label{fig5}
\end{figure}

\section{In which numerical free-energy calculations confirm that the false tiling is thermodynamically stable}
\label{equilibrium}

\subsection{Building the networks} 

To go beyond this simple estimate for rotational free energy we performed numerical free-energy calculations. These allow us to determine the phase behavior of networks, at low pressure, for various values of $w$ and $\phi_0$. We calculated absolute free energies of the network phases, for fixed $N,V,T,w,\phi_0$, using the Frenkel-Ladd method\c{Frenkel-Ladd,Vega}, and calculated free-energy differences using Hamiltonian integration\c{frenkel1996understanding,Vega}.

To build the networks equivalent to the 6.6.6, 4.8.8, and false tilings, we used floppy-box simulations\c{filion2009}. The 4.8.8 and 6.6.6 tilings can be generated almost immediately using this approach. However, finding large unit cells (such as that displayed by the false tiling) is more challenging. We found the false tiling by simulating 40 particles under a range of different pressures and temperatures. Our search was simplified by focusing on fully-bonded networks. Simulation boxes of the resulting structures are shown in \f{boxes}.

We did not find other ordered phases using the floppy-box method, and so the results of this search are consistent with those of dynamic simulation: for the parameter space we considered, both methods yield the 6.6.6, 4.8.8 and false tilings, and no other tilings.
\begin{figure*}
	\begin{tabular}{ccc}
		\includegraphics[width=0.66\columnwidth]{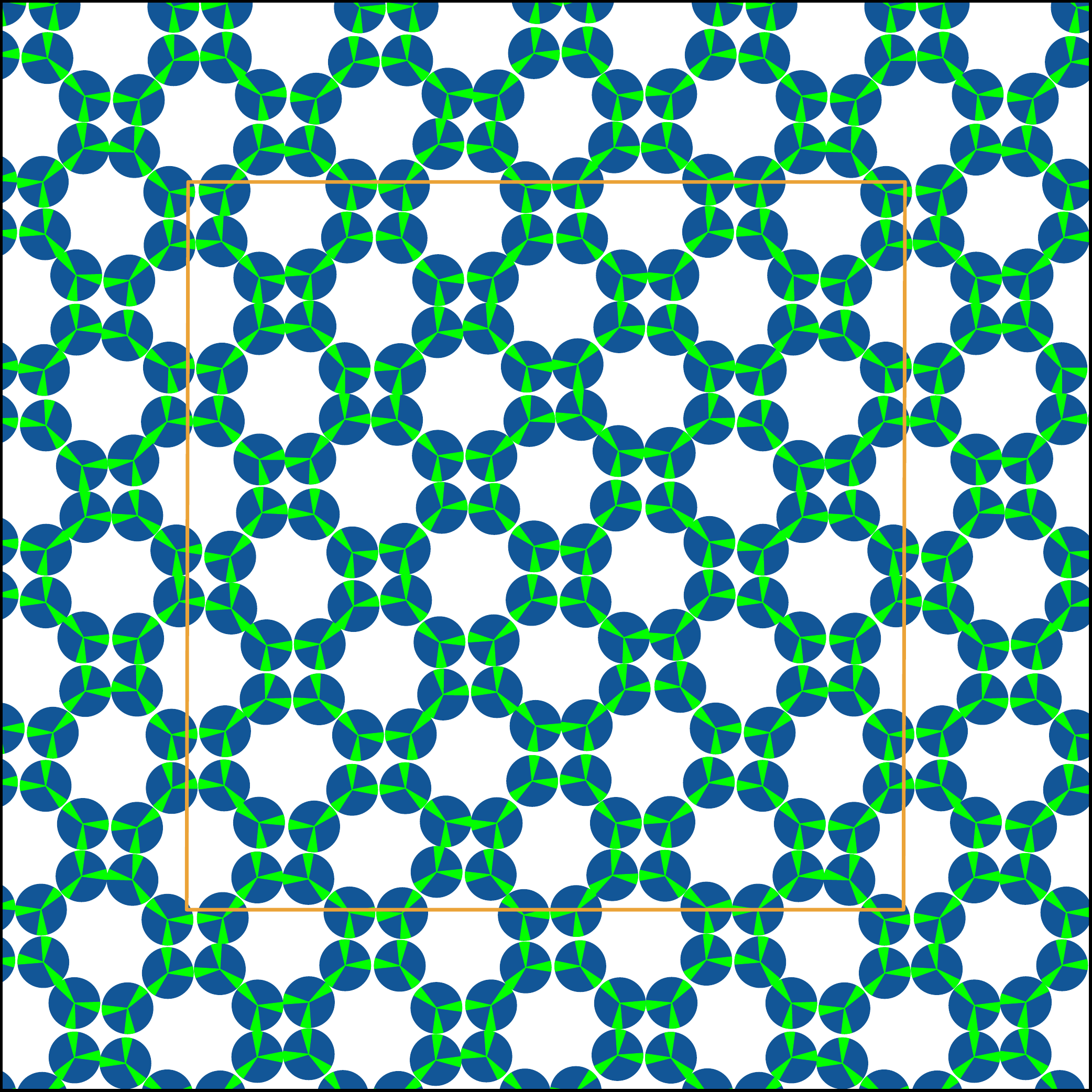} &
		\includegraphics[width=0.66\columnwidth]{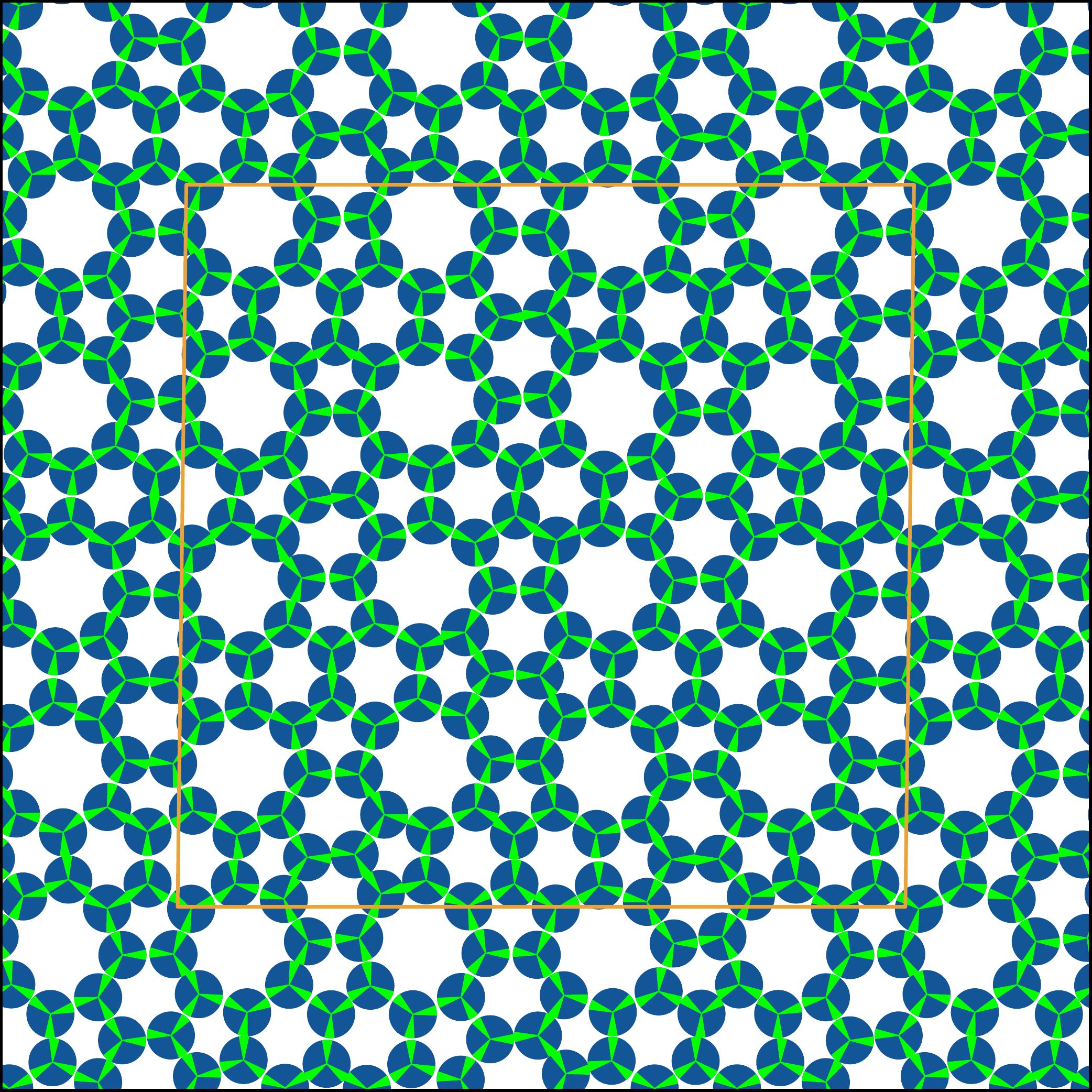} &
		\includegraphics[width=0.66\columnwidth]{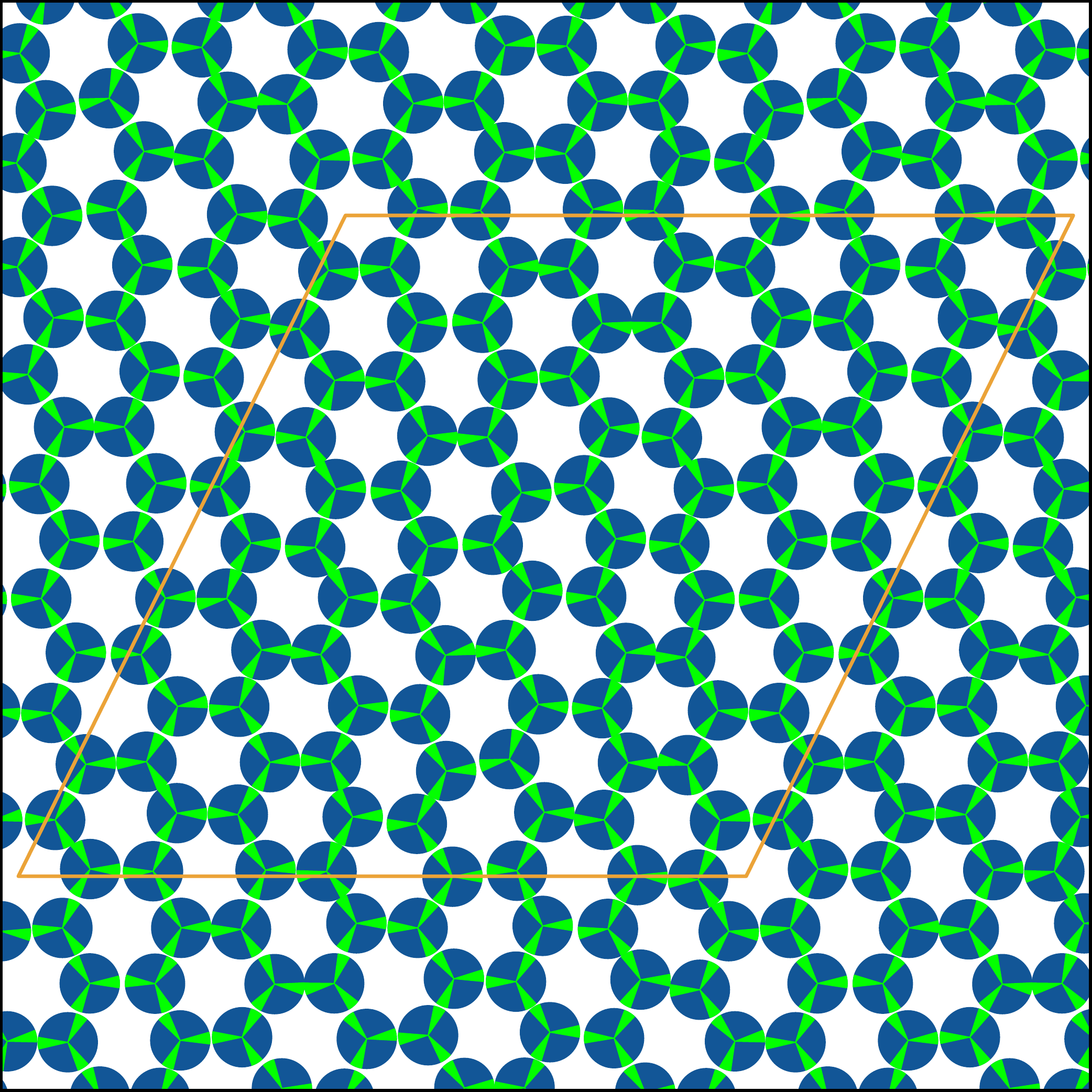} \\
		4.8.8& new & 6.6.6 \\
	\end{tabular}
	\caption{Simulation boxes showing the three tilings. The energy per particle is in all cases $U/N=-3\epsilon/2$.}
	\label{boxes}
\end{figure*}

\subsection{Calculating absolute free energies} 
We performed absolute free-energy calculations of the three phases to confirm their stability and to determine their coexistence properties in the low-temperature, low-pressure limit, which approximates the conditions under which structures in \sec{dynamic} were generated. These calculations closely follow methodology presented in Ref.\c{vissers2013}, using the Kern-Frenkel interaction potential.

We evaluated the absolute free energy of networks using Frenkel-Ladd method, where the reference system is an Einstein crystal with fixed center of mass\c{vissers2013,Vega}.
In the following $f\equiv F/N$ is free energy per particle and $\beta \equiv 1/k_{\rm  B}T$ is the inverse temperature.
The reference translational free energy of an Einstein solid in 2D with fixed center of mass is
\begin{equation}
	\beta f^{\rm  ref}_{\rm trans}=-\dfrac{1}{N}\ln\left[\left(\dfrac{\pi}{\beta\Lambda^{\rm  max}_{\rm trans}}\right)^{N-1}V\right],
\end{equation}
where $\Lambda^{\rm max}_{\rm trans}$ is a large value of the translational coupling, such that all particles are fixed to their lattice positions.

Since specific interactions are present the rotational symmetry of the particles is reduced to that of a one-patch particle in 2D.
For the rotational reference Hamiltonian we use a linear well. The reference rotational free energy than reads
\begin{equation}
	\beta f^{\rm  ref}_{\rm rot}=\ln \int_0^{1}e^{-\beta\Lambda_{r} x}{\rm  d} x=\ln \dfrac{1-e^{\beta\Lambda^{\rm  max}_{\rm r}}}{\beta\Lambda^{\rm  max}_r},
\end{equation}
where $x$ corresponds to the angle between the reference orientation and the particle orientation, and $\Lambda^{\rm  max}_{\rm r}$ is the maximum value of the rotational coupling.
In the cases presented here the energy contribution can be set to a constant, $\beta U=-3N\epsilon/2$. 

\subsection{Calculating free-energy differences}
We wish to identify coexistence lines as a function of the Hamiltonian, and in particular the interaction width $w$ and the angle between patches $\phi_0$. The phase diagram produced is then in terms of these parameters, rather than the traditional pressure-temperature diagrams.

Calculating free-energy differences is usually done using thermodynamic integration or Hamiltonian integration, in which the central object is the derivative $\beta{\partial f}/{\partial \lambda}$.
However, this derivative cannot be directly evaluated for the non-analytic potentials used in this study. We therefore devised an alternative Monte Carlo scheme to calculate the derivatives as a function of $w$ and $\phi_0$. This approach is inspired by previously proposed methods to calculate pressure of particles with hard interactions in the $NVT$ ensemble\c{Vegap}.

\begin{figure*}
	   \includegraphics[width=\linewidth]{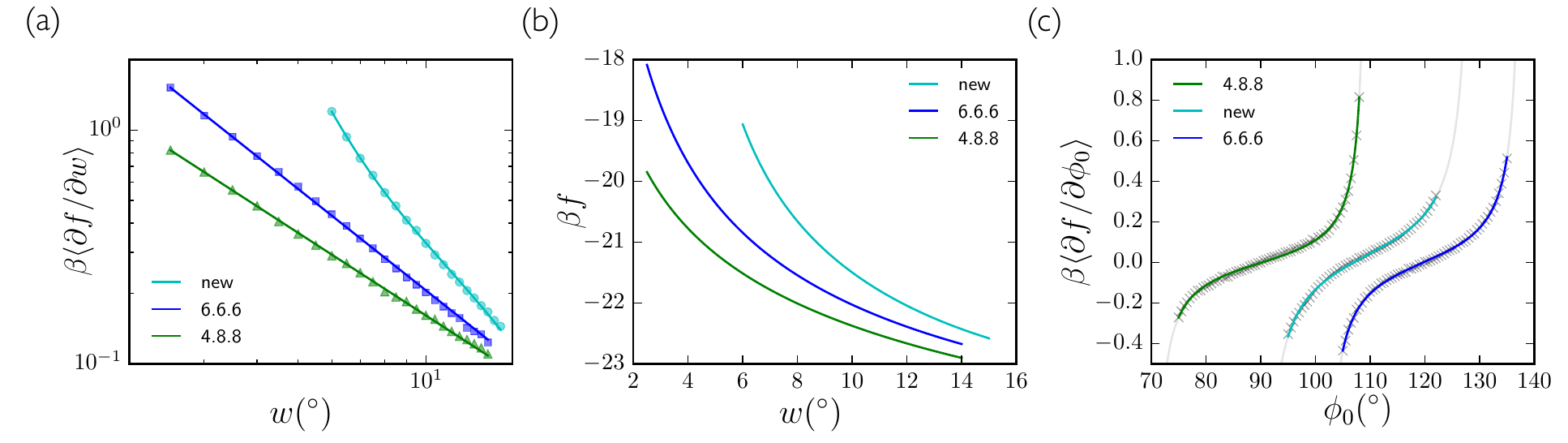} 
\caption{(a) Derivative of the free energy per particle with respect to the patch width $w$, calculated using \eqq{eq:mcdw}. The solid lines are fits. (b) By integrating the data in panel (a), and using absolute free-energy calculations, we obtain the free energies of each tiling as a function of the patch width $w$. The plots for the 6.6.6, 4.8.8 and false tiling (`new') are calculated with $\phi_0=120\deg,90\deg$ and $108\deg$ respectively. (c) Derivative of the free energy as a function of the angle $\phi_0$. The grey crosses are the measured values. The solid lines are the fits using \eqq{eq:tan}. All results are calculated with the patch half-width $w=10\deg$.  }
	\label{fig:dw}
\end{figure*}

We can derive a numerical expression for the required derivative,
\begin{equation}
	-\beta\dfrac{\partial f}{\partial w}=\lim_{\Delta w\to 0^+}\dfrac{1}{\Delta w}\ln\dfrac{Q(N,V,T,w-\Delta w)}{Q(N,V,T,w)},
	\label{eq:mcdw}
\end{equation}
and evaluate the right-hand side of this expression by switching between two canonical ensembles $N,V,T,w_1$ and $N,V,T,w_2$ with an acceptance probability
\begin{equation}
	{\rm acc}(w_1\to w_2)=\min\{1,\exp(-\beta\Delta U)\}.
	\label{eq:acc1}
\end{equation}
The derivative is  the logarithm of the ratio of the number of counts of the states $w=w_1$ and $w_2=w+\Delta w$, i.e. $\beta \partial f/\partial w=\ln P(w_1)/P(w_2)$, where $P(w)$ is the probability that the system is in state $w$.
The derivative as a function of $\phi_0$ is calculated in a similar way, with $w$ replaced by $\phi_0$.
With derivatives in hand the free-energy differences are obtained using integration:
\begin{equation}
	\Delta(\beta f)=\int_{w_1}^{w_2}\av{\beta \dfrac{\partial f(w)}{\partial w}}_{N,V,T,w} \der w.
	\label{eq:tiw}
\end{equation}
The calculated derivatives are plotted in \f{fig:dw}(a) for the 6.6.6, 4.8.8, and false tilings (labeled `new'), taking $\phi_0=120^{\circ}$, $90^{\circ}$ and $108^{\circ}$, respectively. Using \eqq{eq:tiw} in combination with the absolute free-energy calculation yields the free energy of each network as a function of $w$; see \f{fig:dw}(b).

We verified the validity of our free-energy calculation by performing absolute free-energy calculations for multiple values of $w$. For example, the difference in free energies between $w_1=2.5\deg$ and $w_2=14\deg$ for the 6.6.6 tiling, calculated using \eqq{eq:tiw}, is ${\Delta(\beta f)=4.58}$. This value is acceptably close to the difference of the absolute free energies calculated using the Frenkel-Ladd method, $\Delta(\beta f)=\beta f(w_1)-\beta f(w_2)=-18.09+22.69=4.60$.  

\begin{figure}
	   \includegraphics[width=\linewidth]{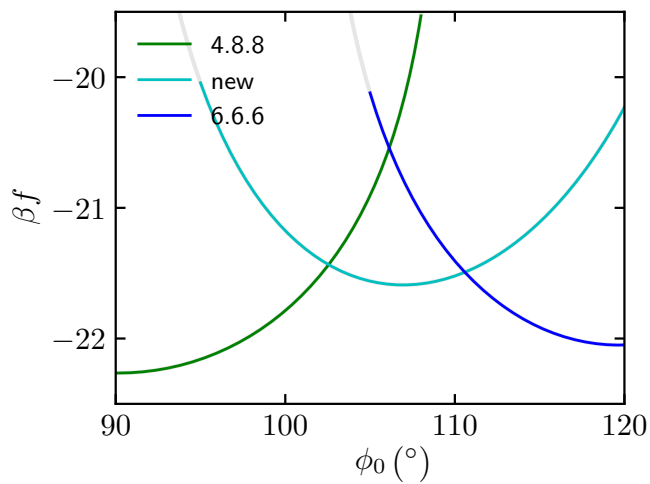} 
\caption{Free energies calculated numerically as a function of the patch angle $\phi_0$ for the 4.8.8, 6.6.6, and false tilings. Temperature is set to $\kt=\epsilon/20$, and the patch width is set to $w=10^{\circ}$ in all cases. These calculations are consistent with the key features of the theoretical prediction shown in \f{fig4}.}
	\label{fig_num}
\end{figure}

Having evaluated the free energies as a function of the patch width, we used the same method to calculate free energies as a function of the angle $\phi_0$.
We plot the derivatives $\av{\beta\partial f/\partial \phi_0}$ in \f{fig:dw}(c) for the patch width $w=10\deg$. The derivatives can be fitted accurately with the function
\begin{equation}
	\av{\beta\partial f/\partial \phi_0}=a\tan[b(\phi_0+\phi^{\rm  min})]+c,
	\label{eq:tan}
\end{equation}
with $a$, $b$ and $c$ constants. This fit facilitates our calculations. $a$ and $b$ govern the slope and the width of the fit, and $\phi^{\rm  min}$ is the angle $\phi_0$ at which the tiling has a minimum free energy. 

\subsection{Free energies and a phase diagram}

Using these methods we obtain the free energy of each network as a function of the angle $\phi_0$: see \f{fig_num}. The coexistence points are where the solid lines intersect. \f{fig_num} can be compared with the simple argument of \sec{analytic}, which yields \f{fig4}. The numerical calculations are in agreement with the key features of the analytic prediction, indicating that the thermodynamics of these tilings is dominated by rotational entropy. The simulation results show, in addition, that some details of tiling thermodynamics (such as the lowest free energies of the 6.6.6 and 4.8.8 tilings) depend upon factors ignored by our simple calculation, such as vibrational entropy.

\begin{figure}
	   \includegraphics[width=\linewidth]{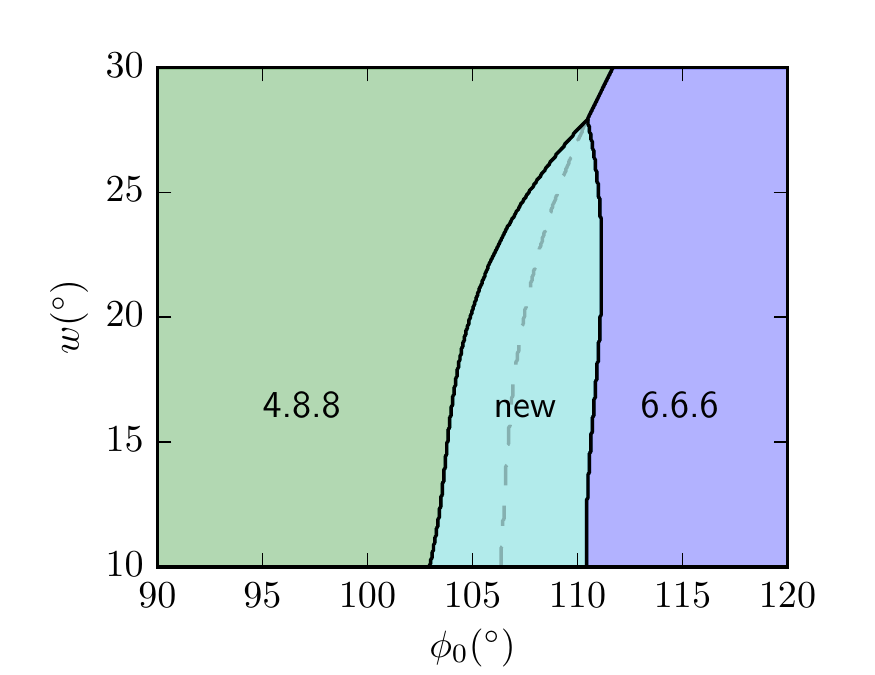} 
 \caption{Low-temperature, low-pressure phase diagram as a function of $w$ and $\phi_0$. The results show that for larger values of the patch width the false tiling becomes metastable with respect to the 6.6.6 and 4.8.8 tilings. The main features of the diagram are consistent with the simple analytic prediction of \f{fig5}. The grey dashed line indicates coexistence between the 6.6.6 and 4.8.8 tilings.}
	\label{fig:pd}
\end{figure}

Having calculated coexistence behavior for the patch width $w=10\deg$, we plot in \f{fig:pd} a low-pressure, low-temperature phase diagram as a function of $w$ and $\phi_0$. To do so, we used the fact that free-energy derivatives are well behaved and can be precisely fit with simple analytical functions, and performed a semi-analytical calculation. This procedure contrasts with the standard method of tracing coexistence lines, in which the Clausius-Clapeyron equation (containing the quantity $\der p/\der T$) is integrated. In principle we could employ a similar strategy because the free energy derivatives with respect to $w$ and $\phi_0$ can be calculated.
But the present method allows us to recover the free-energy landscape over a large parameter space more directly. First, we identified the dependence of the fitting parameters $a,b,c$ and $\phi^{\rm  min}$ (see \eqq{eq:tan}) on $w$, which we plot in \f{fig:fits},
together with the dependence of fitting parameters on $w$; see \f{fig:testfits}. We then evaluated the derivatives \dfdw and \dfdphi at various values of $w$ and $\phi_0$. Both can be evaluated at the same time, and the values of $\Delta w$ and $\Delta\phi_0$ were optimized to maximize the accuracy of the calculation. Finally, we use the fitted solutions to find coexistence points as a function of $w$ and $\phi_0$. 

The phase diagram of \f{fig:pd} is consistent with the key features of the theoretical prediction, which considers only rotational entropy, illustrated in \f{fig5}. In the region of large $w$, where the rotational entropy of phases is similar, simulation shows the 4.8.8 and 6.6.6 tilings to be more stable than the false tiling. The false tiling is thermodynamically stable, or comparable in free energy to the other tilings, in the parameter regime in which we see it in dynamic simulations.

\section{Conclusions}
\label{conclusions}

We have used theory and simulation to show that a model DNA particle self-assembles into a network equivalent to a false tiling. The network is thermodynamically stable, or comparable in free energy to the 6.6.6 and 4.8.8 networks, in the parameter regime in which we see it assemble. The model particle that forms the false tiling has rotational properties that lie between those of the constituents of the 6.6.6 and 4.8.8 tilings. The false tiling is more complex then either of these structures: it possesses three vertex types, none of which is exactly commensurate with the properties of the particle. It is possible that existing DNA nanoparticle designs\c{ANIE:ANIE201601944,doi:10.1021/ja0541938} might be modified to allow the assembly of this structure. More generally, its emergence highlights the fact that unexpected and complex structures can arise from small modifications of regular building blocks.

\begin{acknowledgments}
We thank John Edison for valuable discussions. This work was done as part of a User project at the Molecular Foundry, Lawrence Berkeley National Laboratory, and was supported by the Office of Science, Office of Basic Energy Sciences, of the U.S. Department of Energy under Contract No. DE-AC02--05CH11231
\end{acknowledgments}

%

\clearpage

\onecolumngrid

\renewcommand{\theequation}{A\arabic{equation}}
\renewcommand{\thefigure}{A\arabic{figure}}
\renewcommand{\thesection}{A\arabic{section}}

\setcounter{equation}{0}
\setcounter{section}{0}
\setcounter{figure}{0}

\setlength{\parskip}{0.25cm}%
\setlength{\parindent}{0pt}%

\appendix

\begin{figure}
	\includegraphics{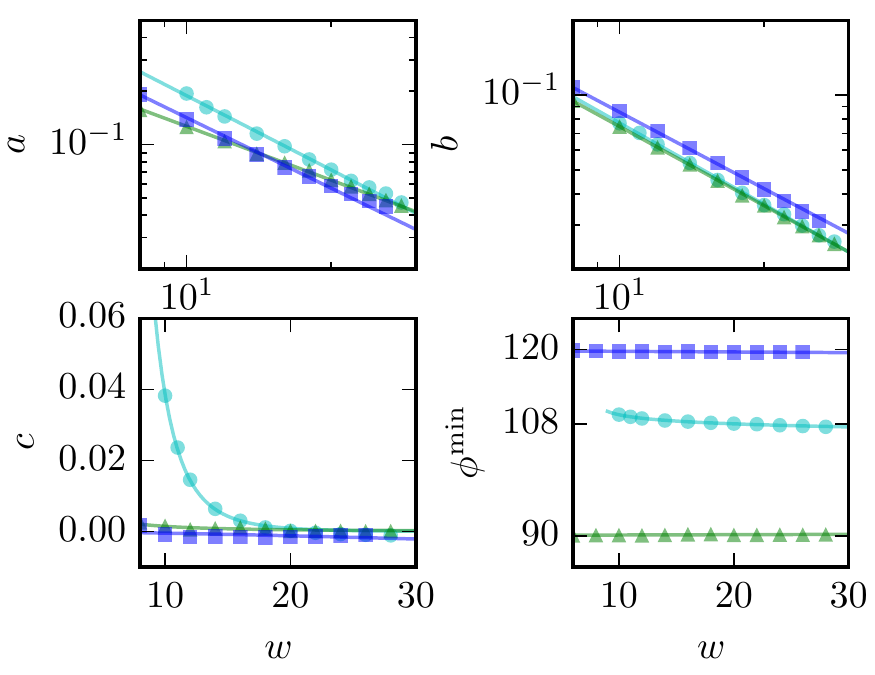}
	\caption{Fitted parameters $a,b,c$ and $\phi^{\rm  min}$ for varying patch width $w$. The blue squares denote the 6.6.6 tiling, the cyan circles denote the false tiling, and the green triangles denote the 4.8.8 tiling. \f{fig:testfits} shows that the parameter-fitting procedure introduces little additional error.}
	\label{fig:fits}
\end{figure}

\begin{figure}
	\includegraphics{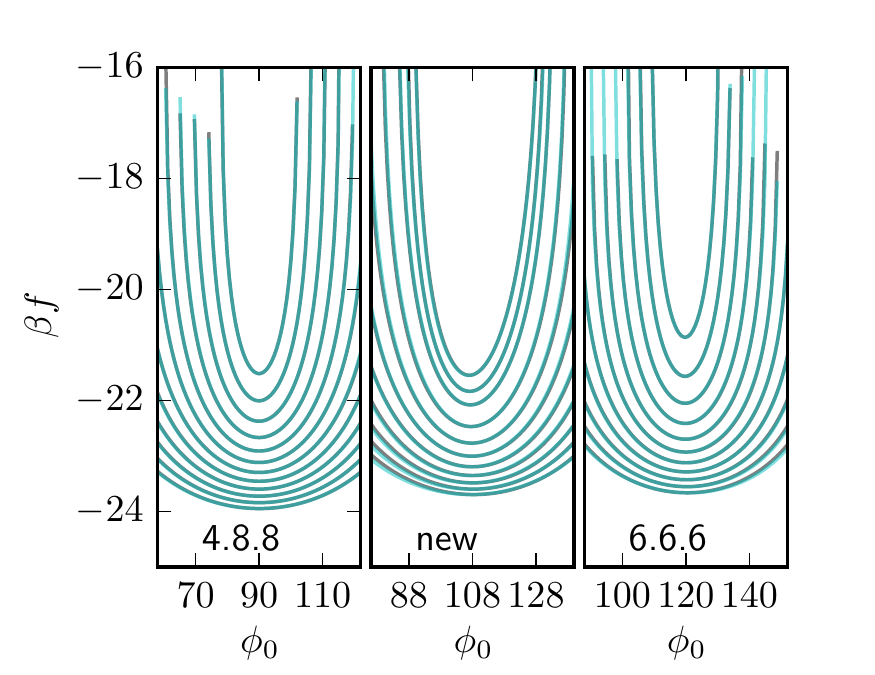}
	\caption{Tiling free energies for various patch widths. The dark green curves are the original numerical calculations (using the parameters $a,b,d,\phi^{\min}$) and the cyan lines are the semi-analytical solutions using the fitted parameters $a,b,c$ and $\phi^{\rm  min}$; see Fig.~\ref{fig:fits}. The dark green and the cyan curves closely match in all cases, indicating that little error is incurred during the parameter-fitting procedure.}
	\label{fig:testfits}
\end{figure}

\end{document}